\documentclass{jpsj-suppl}

\title{Multi-dimensional Core-Collapse Supernova Simulations with Neutrino Transport}

\author{Kuo-Chuan \textsc{Pan}, Matthias \textsc{Liebend\"{o}rfer}, Matthias \textsc{Hempel} and Friedrich-Karl \textsc{Thielemann}}

\inst{Departement Physik,Universit\"{a}t Basel, Klingelbergstrasse 82, CH-4056 Basel, Switzerland}

\email{kuo-chuan.pan@unibas.ch}

\recdate{August 20, 2016}

\abst{

We present multi-dimensional core-collapse supernova simulations using the Isotropic Diffusion Source Approximation (IDSA) 
for the neutrino transport and a modified potential for general relativity in two different supernova codes: 
FLASH and ELEPHANT. 
Due to the complexity of the core-collapse supernova explosion mechanism, 
simulations require not only high-performance computers and the exploitation of GPUs,
but also sophisticated approximations to capture the essential microphysics.
We demonstrate that the IDSA is an elegant and efficient neutrino radiation transfer scheme, 
which is portable to multiple hydrodynamics codes and fast enough to investigate long-term evolutions in two and three dimensions.
Simulations with a 40 solar mass progenitor are presented in both FLASH (1D and 2D) and ELEPHANT (3D) 
as an extreme test condition.
It is found that the black hole formation time is delayed in multiple dimensions and 
we argue that the strong standing accretion shock instability before black hole formation will lead to strong gravitational waves. 
}

\kword{supernovae, neutrinos, black holes, hydrodynamics}

\begin{document}
\maketitle

\section{Introduction}

The Core-Collapse supernova (CCSN) explosion mechanism remains numerically elusive \cite{review1,review2}. 
While it is commonly accepted that neutrino-driven convection and hydrodynamics instabilities are key ingredients
for a successful explosion \cite{summa2016, flashm1, lentz2015,suwa2016,takiwaki2016,skinner2015},
high-resolution, three-dimensional simulations with Boltzmann transport are still numerically too time-consuming.
Therefore, approximated schemes with efficient performance in multiple dimensions are necessary with current computing resources,
but it still has to capture the essential microphysics. 
In this paper, we present multi-dimensional CCSN simulations with the Isotropic Diffusion Source Approximation (IDSA) \cite{idsa} for neutrino transport of electron flavor neutrinos.     

\begin{figure}[tbh]
\centerline{\includegraphics[height=2.in]{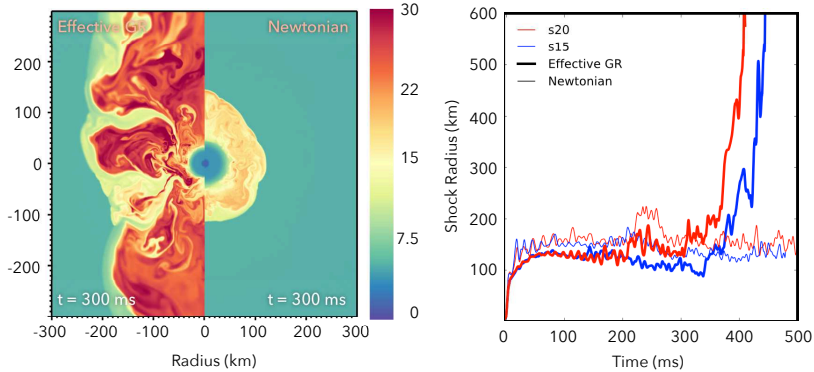}}
\caption{
The left panel shows the entropy distributions of the s20 progenitor from \cite{woosley2007}
at 300 ms postbounce with (left) and without (right) GR correction. 
The right panel indicates the evolutions of average shock radius. 
}
\label{fig1}
\end{figure}

\section{The Isotropic Diffusion Source Approximation}

The IDSA assumes that the distribution function of transported neutrinos can be decomposed 
into a streaming and a trapped component \cite{idsa}. 
The two components are evolved separately and linked by a diffusion source term $\Sigma$. 
To extend the IDSA into multiple dimensions, one could either implement the IDSA 
by a ``Ray-by-ray'' approach in spherical coordinates \cite{suwa2016,takiwaki2016}, 
or solve the diffusion source term in multiple dimensions \cite{idsa2d}.
In this work, we apply the second approach in our two supernova codes. 
Since the trapped particle component could be described by a diffusion problem 
and largely depends on neighbouring zones (see \cite{idsa, idsa2d} for a detailed description), 
we have enabled OpenACC to accelerate the 2D/3D diffusion solver in the IDSA with GPUs.
A speedup of two is currently obtained.  

\section{Supernova codes}

We have ported the IDSA module to both the FISH code \cite{fish} and the publicly available FLASH code \cite{flash, ccsn}. 
We named the merger of FISH with the IDSA ELEPHANT 
(Elegant and Efficient Parallel Hydrodynamics with Approximate Neutrino Transport).
ELEPHANT is a three-dimensional magneto-hydrodynamics code based on an equidistant mesh in Cartesian coordinates. 
The outer layers of the progenitor star are evolved in spherical symmetry by Agile-IDSA \cite{idsa}.    
FLASH is a parallel, multi-dimensional hydrodynamic code based on block-structured Adaptive Mesh Refinement (AMR).
Our current setup supports CCSN simulations in 1D spherical coordinates, 
in 2D cylindrical coordinates, and in 3D Cartesian coordinates. 

\section{General Relativity Potential Correction}

As reported by \cite{liebend2001,muller2012,flashm1}, the higher neutrino luminosities and rms energies 
in (effective) general relativity (GR) calculations favor stronger supernova explosions.  
We have implemented the effective general relativity (GR) potential correction described by \cite{grep} (Case A) 
in both FLASH and ELEPHANT codes. 
Figure~\ref{fig1} gives a comparison of simulations with progenitors s15 and s20 
from \cite{woosley2007} with and without GR correction. 
It is found that the Newtonian cases are failed to explode in 2D while the GR cases explode.

\section{Black Hole Formation in multiple dimensions}

\begin{figure}[tbh]
\centerline{\includegraphics[height=1.38in]{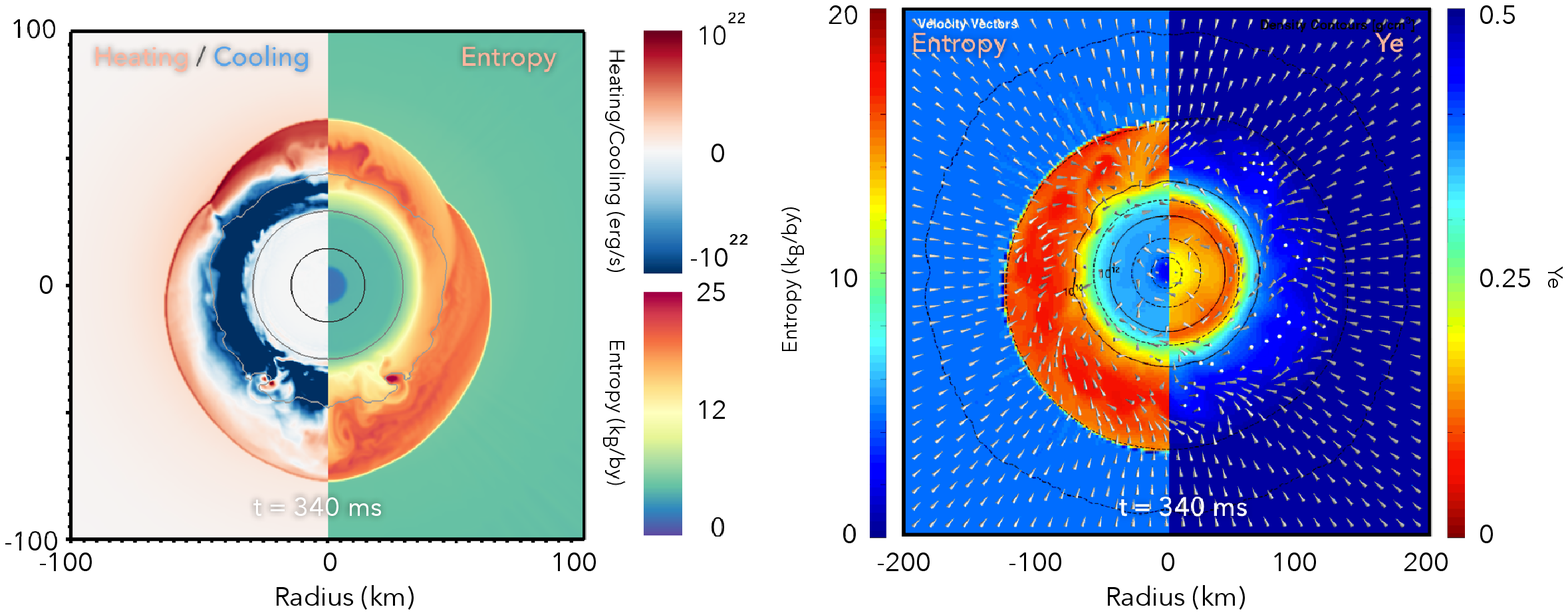}\includegraphics[height=1.4in]{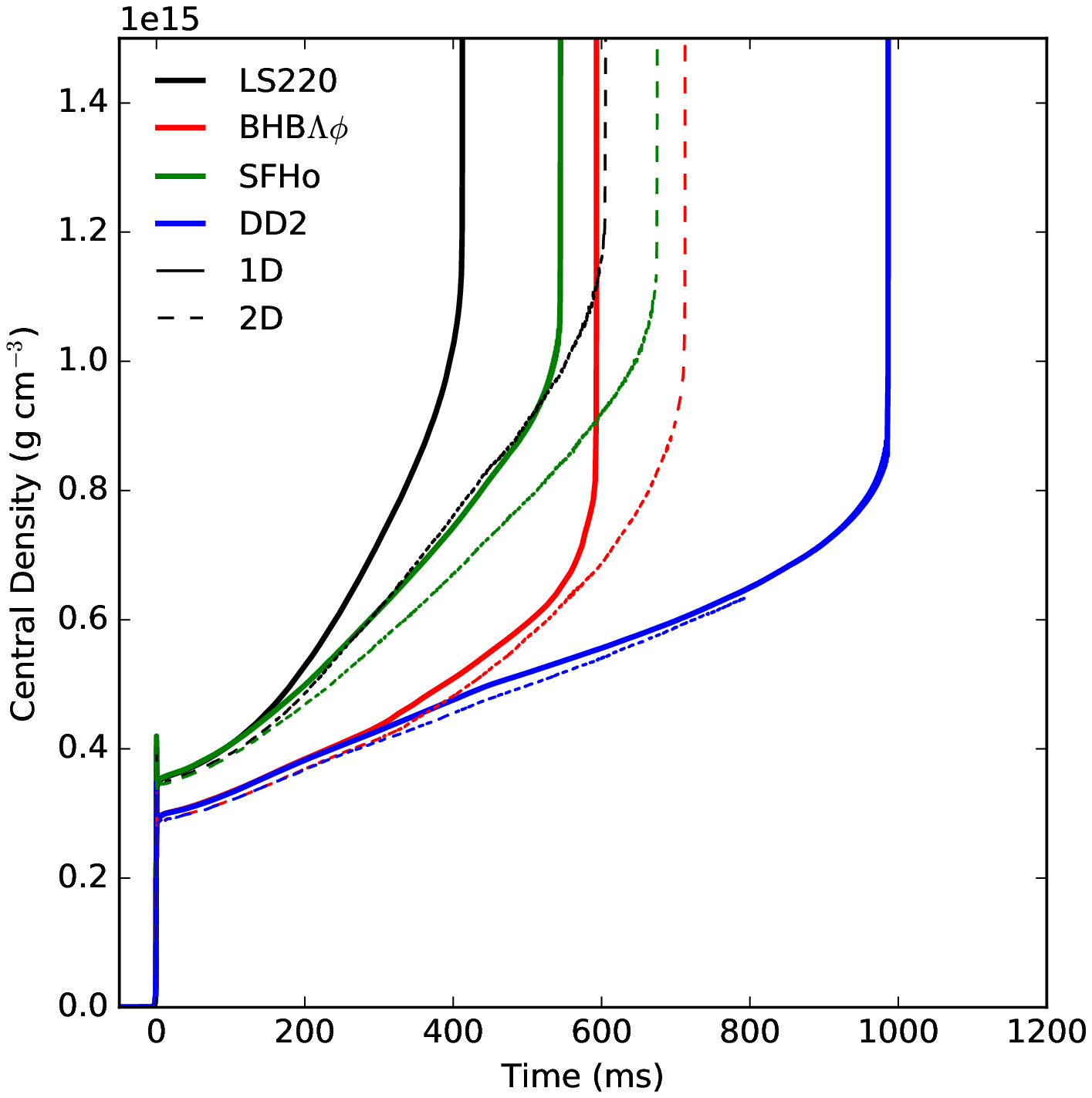} \includegraphics[height=1.45in]{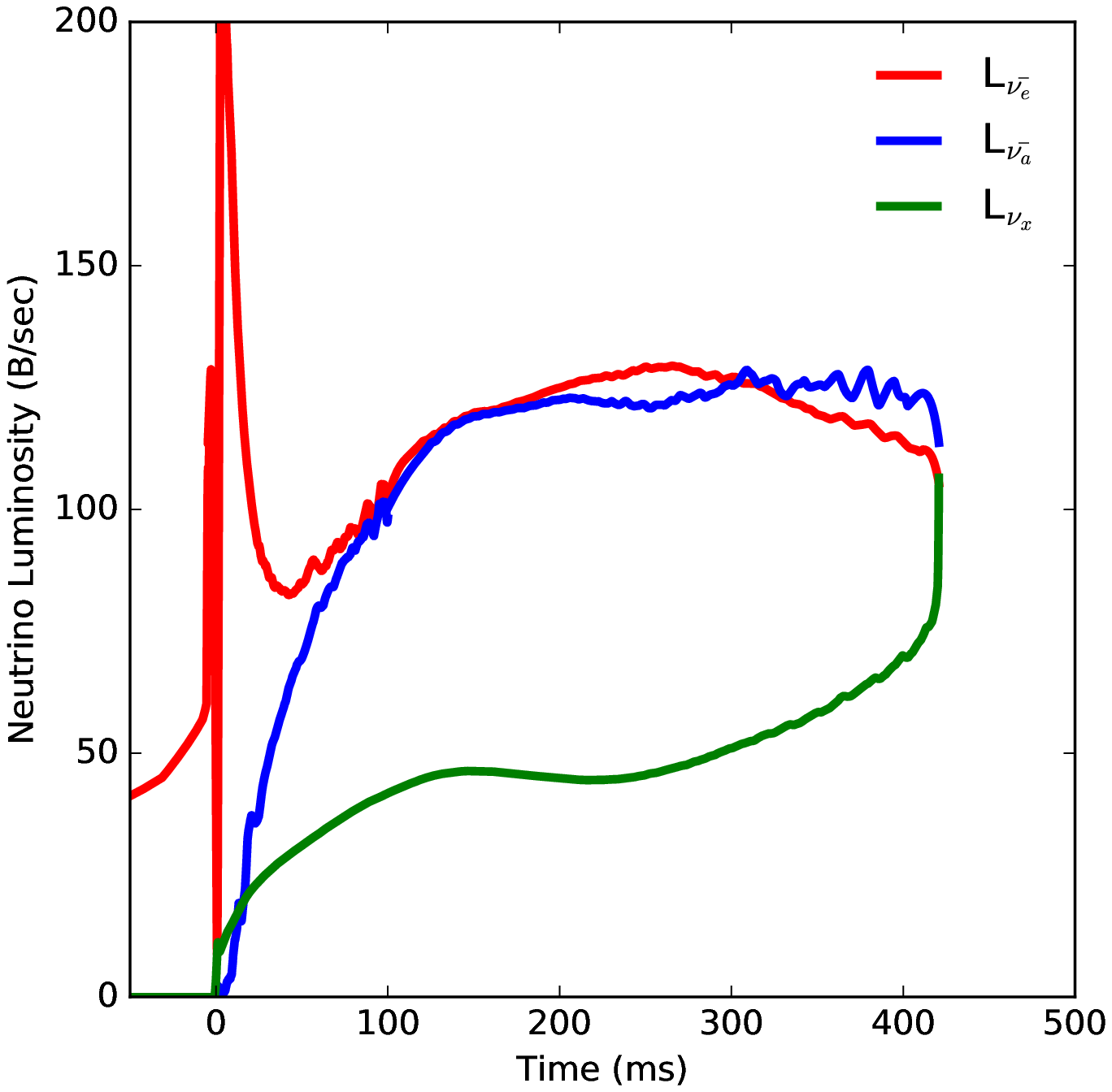}}
\caption{
The left two panels show a comparison of 2D FLASH and 3D ELEPHANT simulations of the s40 progenitor 
with the LS220 \cite{ls220} EoS at 340 ms postbounce. 
The third panel shows the central density of evolutions of the same progenitor with different EoS \cite{ls220,bhblp,sfho,dd2}. 
The right most panel represents a typical neutrino luminosity evolution for a simulation with LS220 EoS.
}
\label{fig2}
\end{figure}

We take the 40 solar mass progenitor, s40, from \cite{woosley2007} to investigate black hole formation
in spherical symmetry and in multiple dimensions with different equations of state (EoS).
Figure~\ref{fig2} shows a comparison of 2D FLASH and 3D ELEPHANT simulations at 340~ms postbounce, 
and a comparison of 1D and 2D FLASH simulations with different EoS \cite{ls220,bhblp,sfho,dd2}. 
A strong standing accretion shock instability (SASI) has been developed in both 2D and 3D cases.   
In addition to the $l=1$ and $l=2$ sloshing modes, $l=1, m=\pm 1$ spiral modes have also been developed in the 3D case.  
The SASI amplitudes are increasing in time while the mass of the proto-neutron star is growing, 
suggesting that strong gravitational waves could be emitted before black hole formation. 
Furthermore, due to the convection and the sloshing motions, the black hole formation time is delayed in multiple dimensions 
and sensitive with respect to the nuclear EoS. 

\section{Conclusions}
We have ported the IDSA module into FLASH and ELEPHANT for neutrino transport and 
have implemented an effective GR potential correction. By performing simulations of a 40 solar mass progenitor as
an extreme condition to study black hole formation in multiple dimensions and with neutrino transport, 
it is found that black hole formation is delayed in multiple dimensions and 
strong gravitational waves may be emitted before black hole formation. 
Although significant work remains, 
we have demonstrated that the IDSA is an elegant and efficient neutrino transport scheme.

  This work was supported by the European Research Council (ERC; FP7) 
under ERC Advanced Grant Agreement N$^\circ$~321263~-~FISH, 
by the PASC High Performance Computing Grant DIAPHANE, 
and by the Swiss National Science Foundation (SNF).
The Basel group is a member of the COST Action New Compstar. 
FLASH was in part developed by the DOE NNSA-ASC OASCR Flash Center at the University of Chicago. 
The simulations have been carried out at the CSCS Piz-Diant under grant No.~661.


\end{document}